\documentclass[preprint,nofootinbib,showpacs,aps,amsmath,showkeys]{revtex4}

\usepackage{cmap}
\usepackage{color}
\usepackage{latexsym,amsmath,amsbsy,graphicx}
\usepackage{float}
\usepackage{footnote}
\usepackage{multirow}
\usepackage{graphicx}
\usepackage{dcolumn}
\usepackage{bm}
\usepackage{color}
\begin{document}   

\title{Double nuclear modification factor in relativistic heavy ion collisions }

\author{A.S.~Chernyshov$^{1}$, I.P.~Lokhtin$^{1}$, A.M.~Snigirev$^{1,2}$ }

\affiliation{$^{1}$ Skobeltsyn Institute of Nuclear Physics, Lomonosov Moscow State University, 119991, Moscow, Russia }

\affiliation{$^{2}$ Bogoliubov Laboratory of Theoretical Physics, Joint Institute for Nuclear Research, 141980, Dubna, Russia }

\begin{abstract}
The hypothesis of the factorization of the double nuclear modification factor in relativistic heavy ion collisions is tested for different types of final-state hadrons within the HYDJET++ model. The results demonstrate that this factorization holds reasonably well at moderately high transverse momenta, provided that the effects of double parton scattering can be small or neglected.
\end{abstract}

\maketitle

\section{Introduction}

In a recent study~\cite{Baranov:2023fez}, a novel observable, the double nuclear modification factor, was proposed as a tool to simultaneously probe initial and final state effects in relativistic nucleus-nucleus collisions. The interplay between double parton scattering and jet quenching effects was illustrated in a simplified scenario where one of the two hard particles propagates through the dense medium without energy loss. In this work, we extend the analysis to a more general case in which both hard particles experience energy loss. We test the hypothesis of factorization of the double nuclear modification factor, originally formulated in~\cite{Baranov:2023fez}.

The paper is organized as follows. Section II outlines the basic definitions and our approach. Section III describes the key features of the HYDJET++ model used to estimate the double nuclear modification factor. Simulation results and their interpretation are presented in Section IV. Finally, our conclusions are summarized in Section V.

\section{Basic definitions}

The experiments at the Relativistic Heavy Ion Collider (RHIC) and the Large Hadron Collider (LHC) have yielded a wealth of information about the properties
of dense and hot nuclear matter (see, e.g., \cite{review} and references therein). One of the most important areas of this research is the search for and study of signals associated with the formation of a new state of matter called quark-gluon plasma (QGP).

A wide range of proposed signals serve as potential indicators of QGP formation. One prominent example is jet quenching in relativistic nucleus-nucleus collisions~\cite{quenching1,quenching2} (see also \cite{review2} for a review of the theoretical framework). Jet quenching probes the mechanisms by which fast partons undergo secondary scattering and energy loss in the dense matter created during the collision. The observable most directly associated with parton energy loss is the suppression of the yield of hadrons with a high transverse momentum $p_{\rm T}$. This suppression is typically quantified by the nuclear modification 
factor, $R_{\rm AA}(p_{\rm T}, \eta)$, which is defined as the ratio of the inclusive single-hadron yield per unit pseudorapidity $\eta$ in $A$-$A$ collisions to the corresponding yield in pp collisions, scaled by the average number of binary nucleon-nucleon ($NN$) collisions $<N_{\rm coll}>$:
\begin{equation} 
R_{\rm AA}(p_{\rm T}, \eta) =\frac{d^2N_{\rm AA}/dp_{\rm T}d\eta}{<N_{\rm coll}>d^2 N_{\rm pp}/dp_{\rm T}d\eta}
=\frac{d^2N_{\rm AA}/dp_{\rm T}d\eta}{<T_{\rm AA}>d^2\sigma_{\rm pp}/dp_{\rm T}d\eta},
\label{eq:quen1}
\end{equation}
where $<N_{\rm coll}>=<T_{\rm AA}> \sigma_{\rm pp}^{\rm inel}$ is the mean number of binary $NN$ collisions in a given centrality class,
$<T_{\rm AA}>$ is the longitudinally integrated nuclear density averaged over the experimentally selected events in a certain collision centrality window, $\sigma_{\rm pp}^{\rm inel}$ is the inelastic non-diffractive pp cross section at a given energy.
$R_{\rm AA}$ quantifies the suppression (or enhancement) of hadron production with respect to $pp$ collisions. If there are no nuclear effects, the value $R_{\rm AA}$
at high enough $p_{\rm T}$ should be unity. 
$R_{\rm AA}$ can be thought of as the survival probability of a particle produced within a nuclear medium. This probability depends not only on energy losses, but also on the spectrum of hard partons produced with respect to their transverse momentum~\cite{Rudolf Baier_2001}. 
For a particle untouched by final state interactions (such as an electroweak boson, $W,Z,\gamma$) we evidently have $R_{\rm AA}=1$ (modulo the effect of nuclear (anti)shadowing on particle distribution functions).

One can generalize this nuclear modification factor to the case of two hard produced  partons (particles) by introducing a double nuclear modification factor~\cite{Baranov:2023fez}:
\begin{equation} 
R_{\rm AA}(p_{\rm T1}, \eta_1;p_{\rm T2}, \eta_2) = \frac{d^4N_{\rm AA}/dp_{\rm T1}d\eta_1 dp_{\rm T2}d\eta_2}{<N_{\rm coll}>d^4 N_{\rm pp}/dp_{\rm T1}d\eta_1dp_{\rm T2}d\eta_2}.
\label{eq:quen2}
\end{equation}

If we assume that the two hard partons do not interact with each other after production and the two final particles are produced through independent hadronization from two separate jets, we can arrive at an expectation that the survival probability will be factorized:

\begin{equation} 
R_{\rm AA}(p_{\rm T1}, \eta_1;p_{\rm T2}, \eta_2)=R_{\rm AA}(p_{\rm T1}, \eta_1) \cdot R_{\rm AA}(p_{\rm T2}, \eta_2).
\label{eq:quen2-1}
\end{equation}

In practice, the assumption of independent survival probabilities for hard partons may not hold. 
In particular, the two final particles may be also  produced from one jet with a violation of the independent hadronization assumption. Besides in single-parton scattering events, both partons are generated at the same spatial point within the nuclear overlap region. Therefore, either they both have a high probability of survival if they are created near the surface of the interaction region, or they both have a low probability of survival when they are created in the inner part of the nuclear overlap.
The particle survival probability may also depend on the kinematic variables as it is noted above.
Then a numerical prefactor is introduced~\cite{Baranov:2023fez} in eq. (\ref{eq:quen2-1}):
\begin{equation} 
R_{\rm AA}(p_{\rm T1}, \eta_1;p_{\rm T2}, \eta_2)=F\cdot R_{\rm AA}(p_{\rm T1}, \eta_1)\cdot R_{\rm AA}(p_{\rm T2}, \eta_2)
\label{eq:quen2+}
\end{equation}
with $F={\cal O}(1)$.

The main purpose of this work is to estimate the prefactor $F$ in a model that deals with jet quenching effect. The HYDJET++ model~\cite{HYDJET_manual} is well suited for this purpose, because the large number of physical observables measured in heavy ion collisions during RHIC and LHC operation are successfully described within this framework. For instance, the model calculations such as transverse momentum spectra, pseudorapidity and centrality dependence of charged particle multiplicity and $\pi^\pm \pi^\pm$ correlation radii~\cite{Lokhtin:2012re}, second and higher-order harmonic coefficients~\cite{Bravina:2013xla}, flow fluctuations~\cite{Bravina:2015sda}, jet quenching observables~\cite{Lokhtin:2011qq,Lokhtin:2014vda}, and angular dihadron correlations~\cite{Eyyubova:2014dha} are in fair agreement with the LHC data for $Pb$+$Pb$ collisions .  

\section{HYDJET++ model}

The HYDJET++ event generator has been developed to simulate relativistic heavy ion $A$-$A$ collisions. It consists of two independent components that describe 
soft and hard processes, respectively.

For soft processes, HYDJET++ employs an adapted version of the FAST MC model~\cite{Amelin:2006qe,Amelin:2007ic}, which is based on a relativistic hydrodynamic parameterization of chemical and thermal freeze-out hypersurfaces under specified freeze-out conditions. The best agreement with experimental data is achieved using a scenario that incorporates separate chemical and thermal freeze-outs. Particle multiplicities are calculated within an effective thermal volume using a statistical model approach. This effective volume incorporates the collective velocity profile and the geometry of the hypersurface, and it cancels out in particle number ratios. Consequently, these ratios remain independent of the freeze-out details, provided that local thermodynamic parameters are spatially uniform. The effective volume concept is used to compute the hadronic composition at both freeze-out stages. The number of particles per event is sampled from a Poisson distribution around a mean value, which is proportional to the number of participating nucleons for a given impact parameter in the $A$-$A$ collision. To simulate elliptic and triangular flow effects, a hydro-inspired parameterization of the momentum and spatial anisotropy of soft hadron emission sources is implemented~\cite{HYDJET_manual,Wiedemann:1997cr}.

For hard processes, the model simulates hard quarks and gluons traversing QGP based on the PYQUEN partonic energy loss model~\cite{Lokhtin:2005px}. This model accounts for both radiative energy loss via gluon emission and collisional energy loss due to parton rescattering in the medium. The number of jets is generated using a binomial distribution, with the mean number per $A$-$A$ event given by the product of the number of binary nucleon-nucleon (NN) sub-collisions (at a given impact parameter) and the integral cross section of the hard process in NN collisions above a minimum transverse momentum threshold, $p_{\rm T}^{\rm min}$. This threshold is a key free parameter of HYDJET++, typically tuned by comparison with experimental data. Partons originating from (semi)hard processes with momentum transfer below $p_{\rm T}^{\rm min}$ are excluded from the hard-parton spectrum; their hadronization products are instead incorporated into the thermalized soft-particle background. Further details of the model are available in the HYDJET++ manual~\cite{HYDJET_manual}, the recent developments and results may be found in~\cite{Chernyshov:2022oik,Ambaryan:2024jhe,Ambaryan:2025}.

\section{Simulation of double nuclear modification factor with HYDJET++ model }

To be specific, we consider $Pb$+$Pb$ collisions for the centrality class $c=0-5\%$ at center-of-mass energy 5.02 TeV per nucleon pair with the free parameters of the model tuned and fixed earlier (in particular, for the successful description of the jet quenching observables~\cite{Lokhtin:2011qq,Lokhtin:2014vda}). The inclusive single and double hadron yields in the central pseudorapidity intervals $|\eta_1|<1$ and $|\eta_2|<1$ are simulated taking into account both jet quenching and nuclear shadowing effects. Table I demonstrates the results on single and double nuclear modification factors for two hadron transverse momentum intervals: $p_{\rm T} > 5 $ GeV/$c$ and $p_{\rm T} > 12 $ GeV/$c$. The deviation $F$ from factorization of the production $\pi$ mesons in association with $K$ mesons and (anti-)protons as well as the production of $K$ mesons in association with (anti-)protons was calculated. The $p_{\rm T}$-dependence of $F$ for the case when both jet quenching and nuclear shadowing are taken into account is shown in Figure 1. It can be seen that the dependences for different types of final-state hadrons are quite similar.

In the case of single nuclear modification factor the effects of jet quenching (Q) and shadowing (S) factorize with a good enough accuracy:
\begin{equation} 
R^{QS}_{\rm AA}(p_{T}, \eta)\simeq R^{Q}_{\rm AA}(p_{T}, \eta)\cdot R^{S}_{\rm AA}(p_{T}, \eta).
\label{eq:quen,s}
\end{equation}

\begin{table}

\caption{\label{tab:table1}
The nuclear modification factors for two different hadron transverse momentum cuts.
 }

\begin{tabular}{cccccccccc} \hline

&$p_{\rm T} >5$ & GeV/$c$ & & $p_{\rm T}>12$ & GeV/$c$ & \\
& & & & & &   \\  

& Q & S& QS & Q& S& QS \\

\hline
    &  & &  &  & & \\      
$R_{AA}^{\pi}$      & 0.34     & 0.71   &0.25 & 0.30& 0.76& 0.24 \\
$R_{AA}^{K}$      & 0.36    & 0.71    &0.26 & 0.31 & 0.79 & 0.23 \\           
$R_{AA}^{p}$      & 0.33    & 0.71    &0.24 & 0.29 & 0.77 & 0.22 \\ \hline


  &  & &  & & &  \\   
$R_{AA}^{\pi K}$      & 0.13     & 0.52   & 0.07 & 0.17 & 0.69& 0.12 \\
$R_{AA}^{\pi p}$      & 0.12     & 0.51    &0.07 & 0.14 & 0.74 &0.12\\
$R_{AA}^{K p}$      &0.12    & 0.51    &0.07  & 0.11 & 0.62 & 0.11\\ \hline

  &  & &  & & &   \\   

$F^{\pi K}$      &1.07     & 1.03    &1.11 &  1.83 & 1.15 & 2.15\\
$F^{\pi p}$      & 1.05     & 1.02    &1.13 & 1.64 & 1.26 & 2.22\\
$F^{K p}$      & 1.03     & 1.02    &1.09 & 1.26 & 1.03 & 2.16 \\
\hline
\end{tabular}
 
\end{table}

\begin{figure}[h]
	\center	\includegraphics[width=35pc]{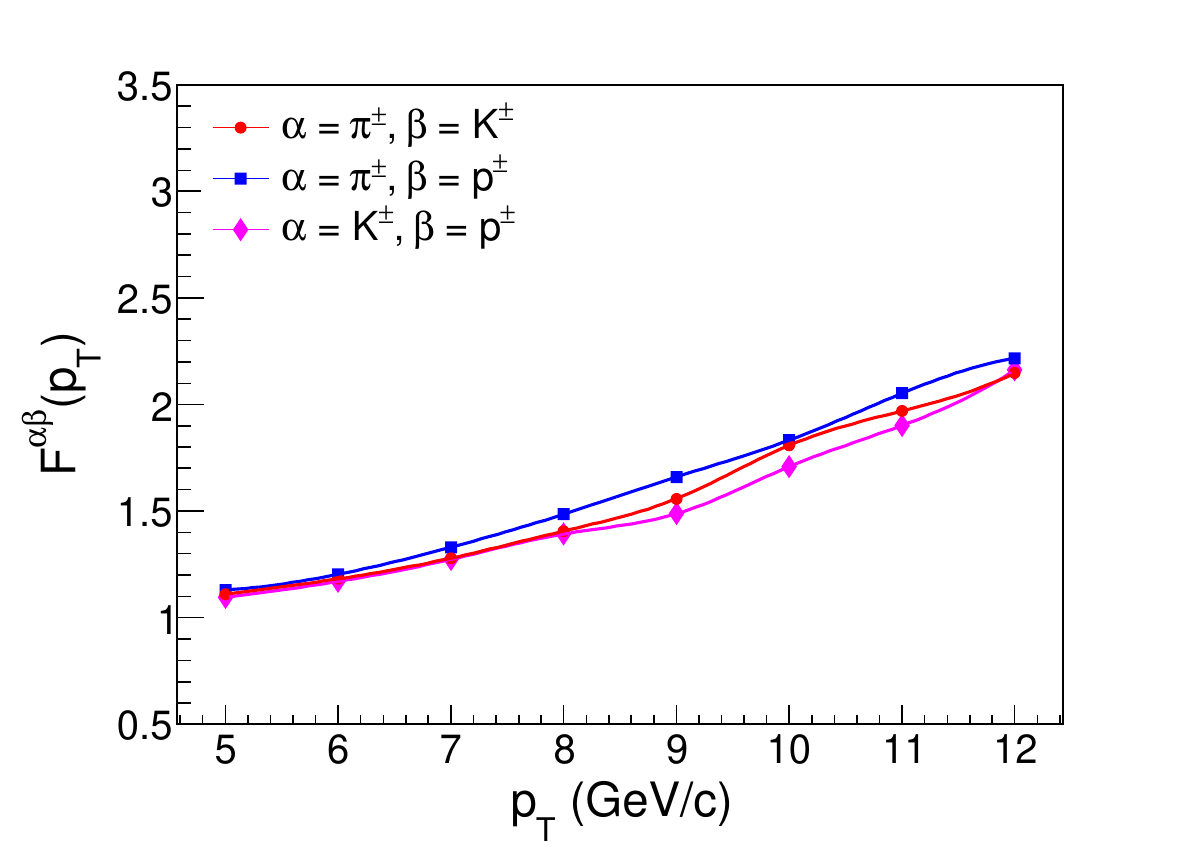}
		\caption{\label{fig1} The dependence of the ratio $F$ of the double and the product of single nuclear modification factors on the minimum value of the hadron transverse momentum $p_{\rm T}$ for mid-rapidity $\mid \eta \mid<1$ in 5\% most central $Pb$+$Pb$ collisions at 
$\sqrt{s_{\rm NN}}=5.02$ TeV (jet quenching and nuclear shadowing are taken into account).}
\end{figure}


As expected, the shadowing correction decreases with increasing $p_{\rm T}$, while jet quenching has a more pronounced impact than shadowing. Here, it should be noted that the modification of nucleon parton distributions is affected not only by nuclear shadowing (suppression at low values of the Bjorken $x$ variable), but also by antishadowing (enhancement at larger $x$). The latter effect is not included in the HYDJET++ model. However, for our purposes, this is not so crucial, as the value of $x\sim 0.1$ (where the maximum antishadowing effect is observed) at a collision energy of 5.02 TeV corresponds to a transverse energy of jets on the order of 250 GeV (for mid-rapidity region). These are relatively rare events, and the dominant contribution for the kinematic domain of particles under consideration is still the shadowing effect.

The deviation $F$ from factorization in the case of a double nuclear modification factor is observable, although it remains relatively small, consistent with the above assertion: 
$F={\cal O}(1)$. The growth of $F$ with increasing $p_{\rm T}$ can be explained by a greater contribution from back-to-back initial hard parton configurations, along with a reduced transverse momentum imbalance. This leads to stronger correlations in the survival probabilities.

\section{Conclusions}

The double nuclear modification factors are studied in central $Pb$+$Pb$ collisions at LHC energy $\sqrt{s_{\rm NN}}=5.02$ TeV for different types of hadrons within the HYDJET++ model framework. Our results indicate that the hypothesis of factorization holds reasonably well at moderately high transverse momenta, provided that the effects of double parton scattering can be small or neglected. However factorization violation starts to show at very high transverse momenta. This supports the conclusion that measurements of the double nuclear modification factors offer significant potential for further investigations. These studies can be extended to various types of hard final-state particles across a broad range of transverse momenta.

\section{Acknowledgments}
A.M.~Snigirev is grateful to S.P.~Baranov, A.V.~Lipatov and M.A.~Malyshev for their previous fruitful discussions.

\section{Funding}
This study was conducted under the state assignment of Lomonosov Moscow State University. A.S.~Chernyshov acknowledges support from the BASIS Foundation under grant No.23-2-10-2-1.

\section{Conflict of interests}
The authors of this work declare that they have no conflicts of interest.

\end{document}